%% file: main.tex
\pdfoutput=1
\documentclass[11pt]{article}

\usepackage{authblk}
\usepackage{ACL2023}
\usepackage{times}
\usepackage{latexsym}
\usepackage[T1]{fontenc}
\usepackage{booktabs}
\usepackage[utf8]{inputenc}
\usepackage{microtype}
\usepackage{inconsolata}
\usepackage{fancyhdr}
\usepackage{graphicx}
\usepackage{caption}
\usepackage{subcaption}
\usepackage{tcolorbox}

\usepackage{amsmath}
\usepackage{amsfonts}
\usepackage{lipsum}

\title{Enhancing Inflation Nowcasting with LLM: \\
    Sentiment Analysis on News}

\author[1]{\textbf{Marc-Antoine Allard}}
\author[1]{\textbf{Paul Teiletche}}
\author[1]{\textbf{Adam Zinebi}}
\affil[1]{EPFL, Lausanne}

\date{}

\begin{document}

\maketitle

\thispagestyle{fancy}
\thispagestyle{empty}

\begin{abstract}
This study explores the integration of large language models (LLMs) into classic inflation nowcasting frameworks, particularly in light of high inflation volatility periods such as the COVID-19 pandemic. We propose \texttt{InflaBERT}, a BERT-based LLM fine-tuned to predict inflation-related sentiment in news. We use this model to produce \texttt{NEWS}, an index capturing the monthly sentiment of the news regarding inflation. Incorporating our expectation index into the Cleveland Fed’s model, which is only based on macroeconomic autoregressive processes, shows a marginal improvement in nowcast accuracy during the pandemic. This highlights the potential of combining sentiment analysis with traditional economic indicators, suggesting further research to refine these methodologies for better real-time inflation monitoring. The source code is available at \href{https://github.com/paultltc/InflaBERT}{https://github.com/paultltc/InflaBERT}.
\end{abstract}

\section{Introduction}
One of the significant economic consequences of the COVID-19 pandemic has been a surge in inflation worldwide. The pandemic caused massive disruptions in global supply chains, a rise in food prices, and the built-up of inflation expectations by economic agents. Inflation is commonly defined as the annualized percent change in consumer prices. According to the World Bank, world inflation surged from 2\% in 2019, before the pandemic outbreak, to 8\% in 2022\footnote{See the World Bank indicator \href{https://data.worldbank.org/indicator/FP.CPI.TOTL.ZG}{here}.}, reaching levels unseen for most countries over nearly three decades. Since then, inflation has receded but remains significantly higher than before the pandemic.
 
Controlling inflation is one of the main objectives of major central banks such as the Federal Reserve of the United States, the European Central Bank, and the Swiss National Bank. Following the inflation shock, these central banks have proceeded with significant interest rate increases that significantly impacted the economy, notably real estate. It is a critical dimension of central banks' work to monitor real-time inflation dynamics. Over the recent years, central banks have notably developed inflation "nowcasting" models. "Nowcasting" stands for the contraction of "Now" and "forecasting". Originating in meteorology, it refers to "the prediction of the very recent past, the present, and the very near future state of an economic indicator" (Wikipedia). In economics, nowcasting has been originally applied to economic growth \cite{banbura2013now}, but models of inflation nowcasting have been recently developed.
 
One such popular model has been developed by Knotek and Zaman \cite{knotek2017nowcasting, knotek2024nowcasting} and is forming the basis of the inflation nowcasting model of the Federal Reserve Bank of Cleveland. The model combines through an Ordinary Least Squares (OLS) regression forecasts of three main inflation components (core inflation, food prices, energy) based on autoregressive processes and high-frequency data such as daily oil prices. The authors show that the forecasts of the nowcasting model outperform those of professional economists (as captured by the Survey of Professional Forecasters – SPF below).
 
More recently, nowcasting models are trying to benefit from machine learning and artificial intelligence development. These methodologies allow researchers to efficiently process a more extensive base of indicators than traditional economic statistics, including internet search trends, social media sentiment, or credit card transaction data, which can all be relevant to tracking the economy and inflation in real-time. Recent examples include Angelico et al. \cite{angelico2022can}, de Bandt et al. \cite{RePEc:bfr:banfra:921}, and Beck et al. \cite{beck2024nowcasting}. The advent of Large Language Models (LLMs) also offers the opportunity to enhance the inflation nowcasting models, as recently shown by Faria e Castro and Leibovici \cite{faria2023artificial} with Google AI's PaLM.
 
This project investigates whether these recent methodologies can improve traditional inflation nowcasting models. We model news inflation sentiment by introducing \texttt{InflaBERT}, a \texttt{BERT}-based~\cite{devlin2018bert} LLM fine-tuned to predict inflation-related news sentiment. To train this model we use an extensive news database extracted from the \texttt{FNSPID} dataset~\cite{dong2024fnspid}. We then propose NEWS, an index replicating monthly news sentiment regarding inflation. Finally, we test whether considering this index can improve the predictions of the Cleveland Fed nowcasting model. We focus on the COVID period (2020-2023) to analyze how these models would have allowed us to better track the strong moves and volatility in inflation during that period.

\section{Related literature}
\label{sec:RL}
Our paper contributes to the literature on improving classic inflation nowcasting techniques through the use of ML and AI techniques to capture expectations. The integration of AI in economic forecasting, specifically inflation prediction, is a burgeoning field that leverages advancements in machine learning and natural language processing. Recent studies, such as the one conducted by Faria-e-Castro and Leibovici (2024) at the Federal Reserve Bank of St. Louis \cite{faria2023artificial}, explore the potential of large language models (LLMs) like Google's PaLM to generate accurate inflation forecasts. This research demonstrates that LLMs can produce conditional inflation forecasts with lower mean-squared errors compared to traditional methods such as the Survey of Professional Forecasters (SPF) over multiple years and forecast horizons. The findings suggest that AI-based models can offer an effective and cost-efficient alternative to traditional expert and survey-based forecasting methods. This study aligns with previous works like those by Bybee \cite{bybee2023surveying}, who used GPT-3.5 to simulate economic expectations, and extends the understanding of how LLMs can be utilized for macroeconomic and financial predictions. 

Despite the promising capabilities of large language models (LLMs) in inflation forecasting, several limitations and challenges persist. Notably, no specific training was conducted on inflation data; instead, instruction tuning was employed. Instructing the LLM to neglect future data is not entirely robust because the model inherently uses weights that were trained on more recent data. This reliance makes it difficult to rigorously test the validity of the model's conditional forecasts as truly out-of-sample. Additionally, the inherent "black-box" nature of LLMs complicates understanding the underlying mechanics driving their predictions, posing challenges for transparency and interpretability in economic forecasting.

\section{News Inflation Sentiment Analysis}
\label{sec:SA}
\begin{figure}[h] 
    \centering 
    \includegraphics[width=\hsize]{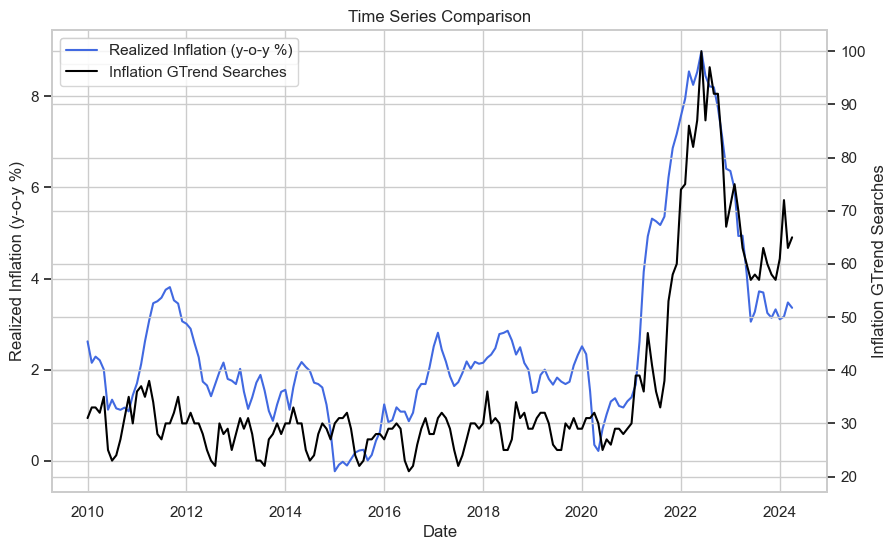} 
    \caption{Number of google searches for ‘inflation’ over time} 
    \label{fig:gtrend} 
\end{figure}

Improving inflation nowcasting by incorporating population expectations seems reasonable when considering the string correlation between human perceptions and realized inflation. For instance, data from sources like Google Trends illustrate this phenomenon, showing that people tend to search more for the term "inflation" during high inflation periods (refer to Figure~\ref{fig:gtrend}).

Recognizing the valuable insights embedded in human behavior and perceptions, we chose news articles as our primary data source. News are comprehensive and highly relevant, reflecting the opinions and signals that traders consider alongside mathematical models and prior information. Lastly, news is readily accessible, which is a significant advantage compared to the costly acquisition of social data, such as Tweets.

\subsection{Data Sources}

\begin{figure}[h] 
    \centering 
    \includegraphics[width=\hsize]{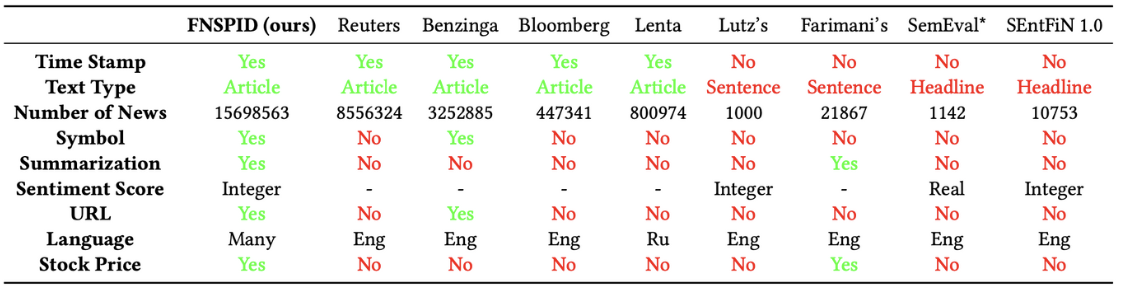} 
    \caption{Comparison of existing datasets for Time Series Financial Analysis} 
    \label{fig:FNSPID} 
\end{figure}

Among the targeted available datasets, we select the Financial News Sentiment and Price Information Dataset (FNSPID). The FNSPID dataset boasts a strong foundation due to its high information content (see figure~\ref{fig:FNSPID}). It contains over 15.7 million time-aligned financial news records spanning 1999 to 2023. Additionally, the dataset is notable as it includes preprocessed article summaries, which condense the text and make it more concise. This reduces the complexity for machine learning models, enabling faster and easier training compared to using long articles with intricate dependencies that are difficult to interpret. 

However, leveraging this valuable data source comes with several challenges. First, processing millions of news articles within our time and resources constraints seems unpractical, imposing down-sampling of the data. Then, this dataset lacks a critical element for training sentiment models: sentiment labels associated with each news article.

\subsection{Data Preparation}
To conduct a sentiment analysis in order to build an effective and expressive Sentiment index, a precise and \textbf{labeled} dataset is needed.

We address these challenges by outlining the preprocessing pipeline we apply to the dataset to obtain training, evaluation, and testing samples. We will also discuss the labeling solution we chose.

\subsubsection*{Preprocessing}
We restrict ourselves to a random sample of 3\% (approximately 500,000 articles) from the FNSPID dataset to address storage and computational constraints. This sample spans the years 1999 to the present day. To facilitate tokenization and expedite model training, we further filter this sample by excluding articles with summaries exceeding 100 words. This results in a substantial dataset of general financial news.

Our next step involves isolating and selecting news specifically related to inflation. We employ a heuristic method based on lexicon selection. This means we only retain articles containing at least one word from our predefined inflation lexicon (which is based on fundamental components of this latest): \\
\textit{{"Inflation", "Gasoline prices", "Food prices", "Deflation", "Consumer price index", "CPI", "Core CPI"}}.\\ This filtering process reduces the dataset from 500,000 articles to over 74,000 articles, while still maintaining a good representation.

Focusing on historical trends of inflation rates, we analyze news articles from 2010 to the present day. This specific timeframe is particularly interesting because it encompasses two distinct phases: a period of low inflation and a period of high inflation. To address the lack of sentiment labels in the dataset, we limit ourselves to a sample of 20 articles per month, creating a manageable dataset for manual labeling.

\subsubsection*{Labeling}

Labeling remains the cornerstone of our project. Inaccurate labels or labels that misrepresent the true sentiment of the news articles in our training set will significantly hinder our ability to develop a high-performing model with strong predictive power. While a simple sentiment classification (positive, negative, or neutral) might seem like a straightforward approach for our initial selection of 2,167 inflation-related news articles, it wouldn't provide the level of granularity we need.

To achieve a more informative labeling system, we have opted for the following classification scheme (details to follow):
\begin{itemize}
    \item \textbf{1} if the article expresses inflation will go up.
    \item \textbf{-1} if the article expresses inflation will go down.
    \item \textbf{0} if the article sentiment about inflation is neutral.
\end{itemize}

While human labeling by financial and economic specialists would have undoubtedly been the ideal approach, budgetary and time constraints inherent to student projects removes this method from our options. However, the recent release of OpenAI's highly performant GPT-4 model has coincided with a surge in Large Language Model (LLM) usage for labeling tasks, evident on platforms like Hugging Face. Recognizing this trend, we have opted to leverage the GPT-4 Turbo API for labeling our news articles using a custom prompt (refer to Appendix~\ref{sec:prompt}).

\begin{figure}[h]
    \centering
        \includegraphics[width=0.8\hsize]{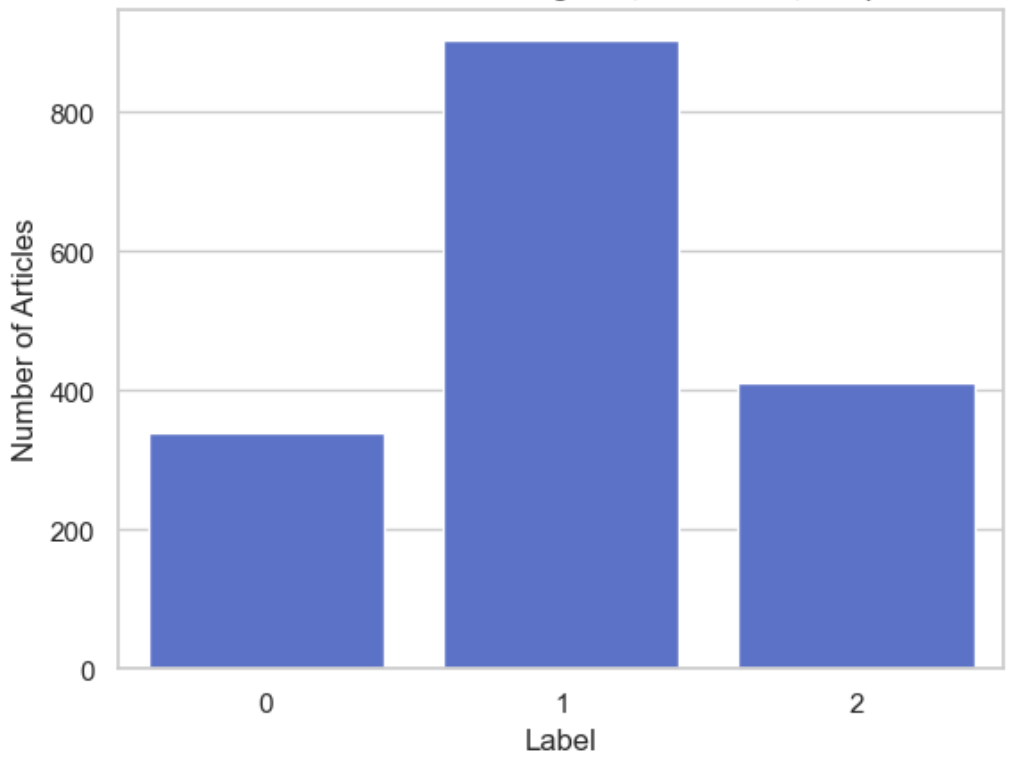}
        \caption{Sentiment Distribution: 0 is negative, 1 neutral, 2 postive}
        \label{fig:DISTRIb}
\end{figure}

The labeling method proves its effectiveness. As shown in Figure~\ref{fig:DISTRIb}, most articles are classified as expressing a neutral sentiment towards inflation. This aligns with our expectations, as many articles likely mention inflation incidentally without explicitly conveying a positive or negative opinion.

However, despite this well-labeled dataset, a challenge persists. Cost constraints prevent further labeling with the expensive GPT-4. Additionally, the class imbalance (predominantly neutral sentiment) combined with the limited dataset size may hinder the effectiveness of fine-tuning a large language model (LLM). There's a high risk that the LLM won't have enough training data to learn the specific nuances of this task. To address the class imbalance, we opt to augment our training set. This process involves increasing the data points in underrepresented classes to achieve a more balanced distribution. We achieve this by leveraging an open-source pre-trained paraphraser, specifically the <\textit{humarin/chatgpt\_paraphraser\_on\_T5\_base}> model~\cite{chatgpt_paraphraser}. This tool allows us to generate paraphrases of existing neutral-sentiment articles, effectively creating synthetic data points for the positive and negative sentiment classes.

Through this data augmentation technique, we expand our training set to a final size of 6,000 labeled news articles.

\subsection{Model}

We conduct a comparative analysis of two prominent machine learning paradigms to identify the most effective models for this sentiment classification task: traditional machine learning models and deep learning models, with a particular focus on transformers, a powerful deep learning architecture.

\subsubsection*{Traditional Machine Learning Techniques}
For each model type, we evaluate two variants: one trained from scratch using Continuous Bag-of-Words (CBOW) or word2vec word embeddings, and another leveraging pre-trained GloVe embeddings (glove.6B.100d).

\begin{itemize}
    \item \textbf{Logistic Regression:} A foundational statistical method widely used for binary classification tasks. While relatively simple, it can achieve strong performance when paired with effective optimization algorithms and regularization techniques (like L1 or L2 regularization).

    \item \textbf{Random Forest:} An ensemble learning method that combines predictions from multiple decision trees. This approach offers several advantages, including interpretability and robustness to high-dimensional and complex data.

    \item \textbf{Support Vector Machine (SVM):} A classic machine learning technique that excels at finding a clear separation between classes in high-dimensional spaces. It can be particularly effective when combined with non-linear kernel functions to handle complex data patterns. However, SVMs can be computationally expensive for very large datasets.

    \item \textbf{XGBoost:} An advanced ensemble learning method based on gradient boosting. It trains multiple decision trees sequentially, where each tree corrects the errors of the previous one. This approach leads to highly accurate models, but XGBoost can be prone to overfitting if not carefully tuned \cite{Chen_2016}.
\end{itemize}

\subsubsection*{Deep Learning Techniques (Transformers Models)}

\begin{itemize}
        \item \textbf{BERT-base:} BERT base version containing 100 million parameters \cite{devlin2018bert}. Its pre-training
        data encompasses the concatenation of the Toronto
        Book Corpus and English Wikipedia. As all others Transformers models BERT use self-attention layer enabling it to encode and align to long term dependencies in the input\cite{vaswani2023attention}.

        \item \textbf{FinBERT:} A pre-trained NLP model to analyze sentiment of financial text \cite{araci2019finbert}. It is built by further training the BERT language model in the finance domain, using a large financial corpus and thereby fine-tuning it for financial sentiment classification. Financial PhraseBank is used for fine-tuning\cite{Malo2014GoodDO}.

        \item \textbf{Fin-distill-RoBERTa:} A distilled version of the RoBERTa-base model \cite{Sanh2019DistilBERTAD}. It follows the same training procedure as DistilBERT. The model has 6 layers, 768 dimension and 12 heads, totalizing 82M parameters. It was also fine-tuned on the on the Financial Phrasebank dataset\cite{Malo2014GoodDO}.
        
    \end{itemize}

\subsection{Experiment}
For the selection and training procedure, we first split our dataset into training and testing sets based on timeframe rather than a fixed ratio. The training split contains news articles from January 1st, 2010 to December 31st, 2019 (low inflation period), and the test split contains articles from January 1st, 2020 to December 31st, 2023 (high inflation period). As mentioned earlier, out-of-sample testing is crucial for our case.

We begin by selecting the best performing configuration from the traditional machine learning models (refer to~Appendix~\ref{sec:cv-res} to see all the hyperparameters values tested) . We achieve this through a grid search with cross-validation using a 3-fold k-fold procedure. A train-evaluation split with a 0.1 ratio is applied to the initial training set.

Next, using the same split, we fine-tune our transformer models to assess and improve their effectiveness in sentiment analysis tasks. We employ the Transformers library~\cite{wolf-etal-2020-transformers} to fine-tune the three large language models (LLMs) for our specific task. This process involves conducting 5 training epochs. The choice of five epochs is a balance between achieving realistic training within a reasonable timeframe and considering resource constraints. Given the relatively small dataset size, employing more epochs could lead to overfitting. For this benchmarking step, we utilize the AdamW optimizer with a fixed learning rate of 2e-5 and a batch size of 16.

\subsection{Results}
The benchmarking of the traditional machine learning models highlights the superiority to XGBoost in a similar and rigid test setup (see Table~\ref{table:results_trad}).

\begin{figure}[h]
    \centering
    \begin{subfigure}[b]{\hsize}
        \includegraphics[width=\textwidth]{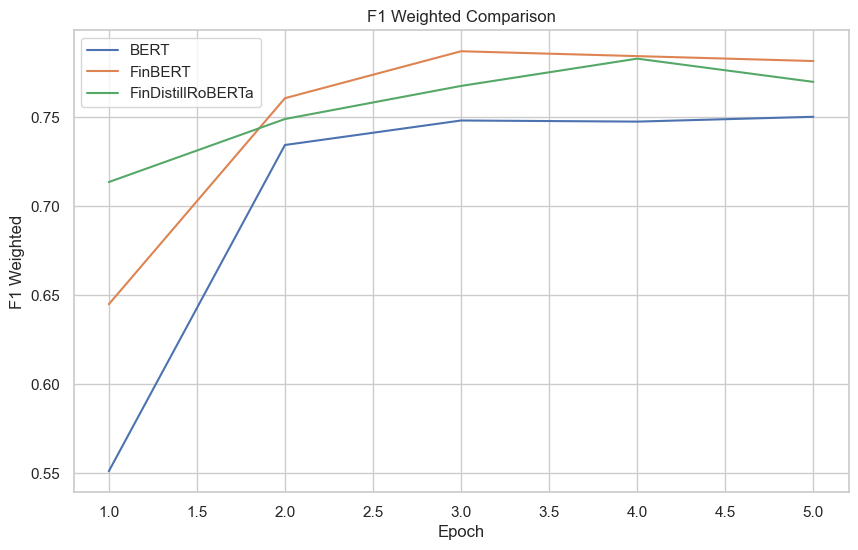}
        \caption{Evaluation F1-weighted}
        \label{fig:f1eval}
    \end{subfigure}
    \hfill
    \begin{subfigure}[b]{\hsize}
        \includegraphics[width=\textwidth]{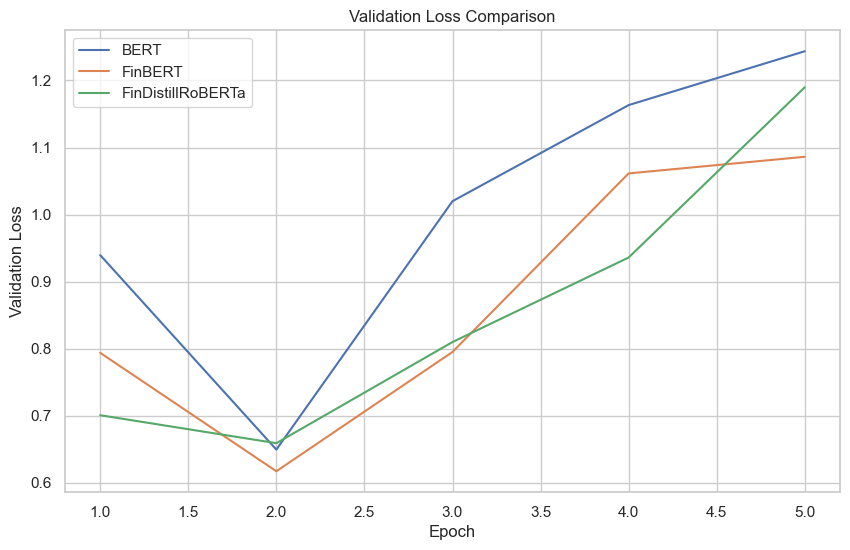}
        \caption{Evaluation loss}
        \label{fig:evalloss}
    \end{subfigure}
    \caption{Training Analytics: Transformers Models}
    \label{fig:analytics}
\end{figure}

For the transformer models, we identify the best epoch checkpoints based on the evaluation loss during the fine-tuning process (see Figure~\ref{fig:analytics}). Subsequently, we use these checkpoint models on the test set alongside the two best XGBoost configurations to determine the overall best performing model (see results Table~\ref{table:results_global}).

\begin{table*}[h!]
    \centering
    \begin{tabular}{lcccc}
        \toprule
        Model & F1 Label 0 & F1 Label 1 & F1 Label 2 & F1 Weighted \\
        \midrule
        Fin-distill-RoBERTa & \textbf{0.486842} & 0.527132 & \textbf{0.677419} & \textbf{0.570034} \\
        FinBERT    & 0.378378 & 0.522059 & 0.586667 & 0.516634 \\
        BERT-case       & 0.416000 & \textbf{0.560261} & 0.625000 & 0.554771 \\
        \midrule
        XGBoost W2V & 0.449869 & 0.111111 & 0.645598 & 0.451693 \\
        XGBoost GloVe & 0.487519 & 0.139130 & 0.615819 & 0.487519 \\
        \bottomrule
    \end{tabular}
    \caption{F1 Scores on the Test Set (Out Sample)}
    \label{table:results_global}
\end{table*}

As mentioned earlier, XGBoost is powerful, but both versions exhibit overfitting on strongly positive and negative sentiments (see Figure X for details, where applicable). However, in terms of overall performance on detecting all three sentiment classes, the best performing model is Fin-distill-RoBERTa, achieving a strong F1-weighted score of \textbf{0.570034} on the out-of-sample test set.

This new high-performing fine-tuned sentiment analysis model, named InflaBERT, will be the model used in our research. InflaBERT can be found on our Hugging Face page: \href{https://huggingface.co/MAPAi/InflaBERT}{https://huggingface.co/MAPAi/InflaBERT}.

\section{Inflation Nowcaster}
\label{sec:nowcast}
\input{nowcaster.tex}

\section{Conclusion}
\label{sec:conclusion}
The ongoing development of AI and ML technologies holds significant promise for more accurate and efficient economic nowcasting, thereby aiding central banks and policymakers in their critical decision-making processes. In particular, the direct use of LLMs demonstrates promising advancements in economic forecasting methodologies (as shown by Faria-e Castro~\cite{faria2023artificial} and Bybee~\cite{bybee2023surveying}). This work, through the development of InflaBERT—a BERT-based model fine-tuned for predicting inflation-related news sentiment—has shown that leveraging advanced machine learning techniques can enhance traditional models like the Cleveland Fed's nowcasting model. The creation of the NEWS index from InflaBERT’s sentiment analysis offered a novel dimension to nowcasting, particularly during the volatile economic period of COVID-19.

The results indicated that the inclusion of the NEWS index provided a marginal improvement in forecast accuracy, as measured by a slight reduction in RMSE. Although the improvement was not statistically significant, it suggests potential for further refinement. This indicates that machine learning and sentiment analysis can complement traditional economic models, providing a more nuanced understanding of real-time inflation dynamics.

Future research could aim to address the observed limitations by improving the accuracy of the \texttt{NEWS} sentiment index through a larger labeled dataset and utilizing more advanced models, such as the latest open-source generative models. Additionally, exploring non-linear nowcasters, such as LSTM models which have demonstrated stronger performance in similar tasks~\cite{siami2018comparison, cao2019financial}, could better capture the complexities of economic indicators.


\clearpage
\bibliography{anthology,custom}
\bibliographystyle{acl_natbib}


\clearpage
\section*{Appendix}
\appendix
\input{appendix.tex}

\end{document}

%% file: nowcaster.tex
\subsection{NEWS Index}
Using \texttt{InflaBERT} (derived in section~\ref{sec:SA}), we construct an index replicating the news sentiment regarding inflation.

\begin{figure}[h]
    \centering
    \begin{subfigure}[b]{\hsize}
        \includegraphics[width=\textwidth]{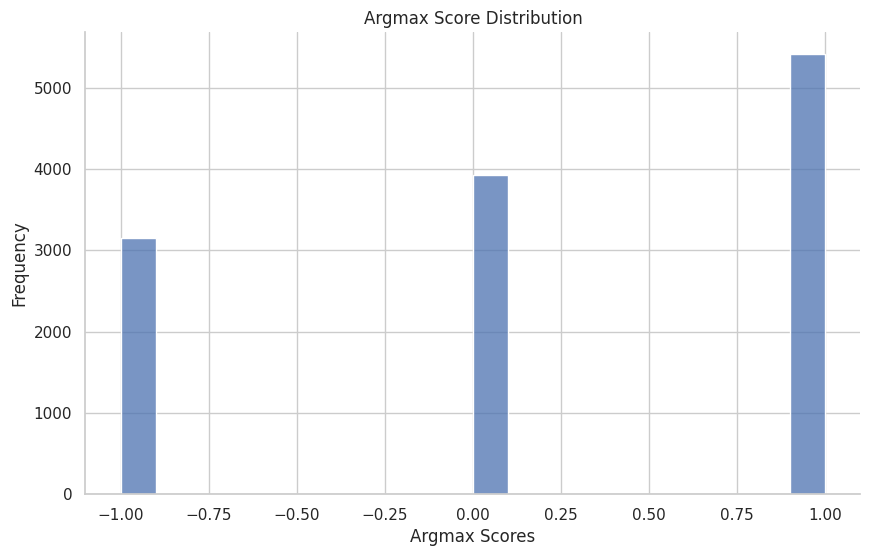}
        \caption{Argmax Score Frequencies}
        \label{fig:argmax}
    \end{subfigure}
    \hfill
    \begin{subfigure}[b]{\hsize}
        \includegraphics[width=\textwidth]{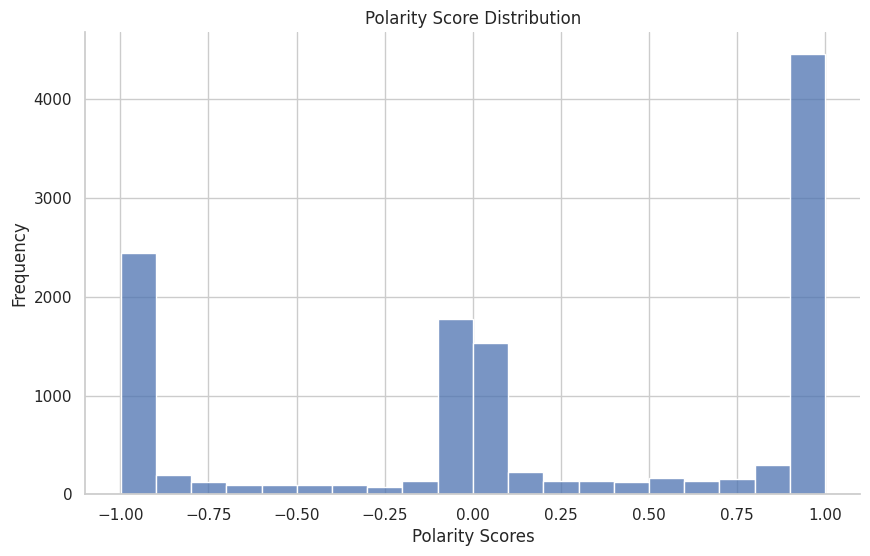}
        \caption{Polarity Score Frequencies}
        \label{fig:polarity}
    \end{subfigure}
    \caption{Different Scores Frequencies}
    \label{fig:scores}
\end{figure}

To accomplish this, we first need to select a score function that transforms our sentiment probabilities into individual scores, which we can then aggregate for each news item within a period. Initially, we consider the argmax score -- $score(n) = \text{argmax}\{\text{InflaBERT}(n)\}$ -- which is a straightforward and intuitive method that assigns scores based on the most probable label. However, this method is highly sharp (see figure~\ref{fig:argmax}) and does not account for the model's uncertainty. Therefore, we also consider an alternative, the polarity score:

$$
score(n) = \mathbb{E}[\text{InflaBERT}(n)]
$$

Where $\mathbb{E}[\text{InflaBERT}(n)]$ is the model's expectation regarding a news $n$ vis-à-vis its labels $l$, i.e:

$$
\mathbb{E}[\text{InflaBERT}(n)]=\sum_{l}{l \cdot p_l^{\text{InflaBERT}}(n)}
$$

The polarity score has a significant advantage as it accounts for the model's uncertainty. When \texttt{InflaBERT} is uncertain between labels, it produces a weighted average of the labels based on their respective certainties. This results in a smoother scoring function (see Figure~\ref{fig:polarity}).

Then, we apply our scoring function to each news of our dataset and aggregate them monthly using cumulative expectation:

$$
P^{\text{NEWS}}_t = \sum_{i=0}^{t}\mathbb{E}_{N_i}[score(N_i)]
$$

In practice, we don't have access to the exact monthly news inflation sentiment -- $\mathbb{E}_{N_i}[polarity(N_i)]$ --, so we estimate it using the sample mean, that we know from the strong Law of Large Numbers~\cite{chung2008strong} converge asymptotically to the population monthly news inflation sentiment: 
$$
\frac{1}{|N_i|}\sum_{n \in N_i}{score(n)}
$$

\begin{figure}[h] 
    \centering 
    \includegraphics[width=\hsize]{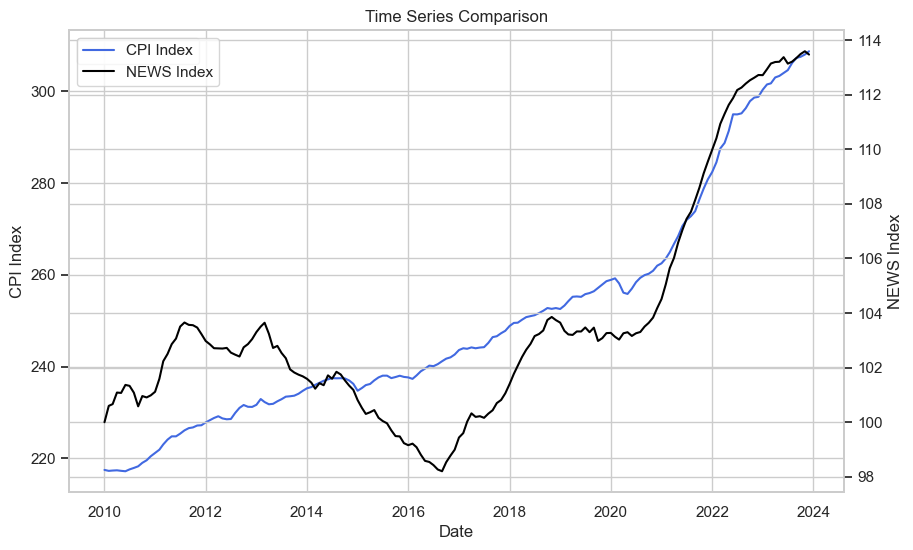} 
    \caption{The \texttt{NEWS} index over the 2010-2023 period} 
    \label{fig:NEWS} 
\end{figure}

This leads to \texttt{NEWS}, an index replicating the news sentiment regarding inflation, as shown in figure~\ref{fig:NEWS}.

Even though the \texttt{NEWS} index captures the high inflation period during COVID-19 quite well, it appears overly aggressive (predicting periods of high deflation and high inflation) during stable periods of realized inflation. This indicates that further work is needed to stabilize the index, such as smoothing the scoring function even more, training on more data, or using a more complex model.
 
\subsection{NEWS For Inflation Nowcasting}
We aim to enhance state-of-the-art inflation nowcasters by incorporating our \texttt{NEWS} index. To achieve this, we modify the Cleveland Federal Reserve Bank model~\cite{knotek2017nowcasting} by adding a term that reflects the influence of our \texttt{NEWS} index.

The Cleveland model is an OLS regression that relates the percentage changes in main inflation components -- Core CPI, Food CPI, and seasonally adjusted Gasoline CPI -- to overall inflation. Specifically, let $\pi_t^{\text{i}} = 100*(P^{\text{i}}_t/P^{\text{i}}_{t-12}-1)$ represent the percentage change of index $i$, the Cleveland model is then defined as follows:

\begin{equation}
\label{eqn:fed}
    \resizebox{.85\hsize}{!}{
        $\pi_t^{\text{CPI}, \text{FED}} = \beta_0 + \beta_1 \pi_t^{\text{Core CPI}} + \beta_2 \pi_t^{\text{Food CPI}} + \beta_3 \pi_t^{\text{Gasoline}} + e_t$
    }
\end{equation}

Adding the \texttt{NEWS} index to the Cleveland model described in (\ref{eqn:fed}), we obtain our new model (see model~(\ref{eqn:ours})). 


\begin{equation}
    \label{eqn:ours}
    \resizebox{.85\hsize}{!}{
        $\pi_t^{\text{CPI}, \text{FED+NEWS}} = \pi_t^{\text{CPI}, \text{FED}} + \beta_4 \pi_t^{\text{NEWS}} + e'_t$
    }
\end{equation}

\subsubsection{Experiment}
To assess the enhancement our novel index can provide, we compare our two models from COVID-19 (January 2020 to December 2023). This analysis help to determine how effectively the new index could have tracked the movements in inflation during that highly volatile time. Notably, \texttt{InflaBERT} is out-of-sample for this period, ensuring that our news sentiment index remains unbiased by previously observed news.

The models are trained on data from January 2015 to December 2019, covering the 60 preceding periods as suggested by the baseline authors~\cite{knotek2017nowcasting}. 

To compute the prediction for a given month, we need the values of all the indices at that time. However, the values for Core CPI, Food CPI, and Gasoline for month $t$ are released only around the 15th day of the following month ($t+1$). Therefore, we replace the exact values in the nowcast with a prediction obtained by computing the moving average of the indices over the past year (12 periods), i.e: $\hat{\pi}_t^{\text{i}} = \frac{1}{12}\sum_{k=1}^{12}{\pi_{t-k}^{\text{i}}}$. The prediction of inflation at a month $t$ can thus be computed around the 15th day of the month $t$ and is formulated as follows:

$$
\hat{\pi}_t^{\text{CPI}} = \beta_0 + \beta_1 \hat{\pi}_t^{\text{Core CPI}} + \beta_2 \hat{\pi}_t^{\text{Food CPI}} + \beta_3 \hat{\pi}_t^{\text{Gasoline}}
$$

For the model that includes the \texttt{NEWS} index, assuming we have access to the news from the first 15 days of month $t$, we can compute the exact value of $\pi_{t}^{\text{NEWS}}$ and use it to nowcast.

We evaluate our nowcaster models on annualized month-over-month inflation predictions $\hat{\pi}_t^{\text{CPI}, \text{Annualized}} = 100*[(\frac{\hat{\pi}_t^{\text{CPI}}}{100}+1)^{12}-1]$. We quantify the significance of our improvements by employing a methodology ~\cite{knotek2023real} using Root Mean Square Error (RMSE) and the Giacomini-White test~\cite{giacomini2006tests}.

\subsubsection{Results}

Table~\ref{tab:reg_res} presents the OLS regression results, indicating that the coefficient related to the \texttt{NEWS} index is statistically significant at the 10\% level.

\begin{table}[!htbp] \centering
\begin{tabular}{@{\extracolsep{5pt}}lcc}
\\[-1.8ex]\hline
\hline \\[-1.8ex]
& \multicolumn{2}{c}{\textit{Dependent variable: CPI}} \
\cr \cline{2-3}
\\[-1.8ex] & \multicolumn{1}{c}{fed} & \multicolumn{1}{c}{fed+news}  \\
\\[-1.8ex] & (1) & (2) \\
\hline \\[-1.8ex]
 const & 0.021$^{}$ & 0.017$^{}$ \\
& (0.026) & (0.025) \\
 pi-CCPI & 0.616$^{***}$ & 0.634$^{***}$ \\
& (0.135) & (0.133) \\
 pi-FCPI & 0.186$^{***}$ & 0.176$^{***}$ \\
& (0.064) & (0.063) \\
 pi-Gasoline & 0.035$^{***}$ & 0.034$^{***}$ \\
& (0.002) & (0.002) \\
 pi-NEWS & & 0.149$^{*}$ \\
& & (0.081) \\
\hline \\[-1.8ex]
 Observations & 60 & 60 \\
 $R^2$ & 0.892 & 0.898 \\
 F Statistic & 153.612$^{***}$ & 120.877$^{***}$ \\
\hline
\hline \\[-1.8ex]
\textit{Note:} & \multicolumn{2}{r}{$^{*}$p$<$0.1; $^{**}$p$<$0.05; $^{***}$p$<$0.01} \\
\end{tabular}
\caption{Regression Results}
\label{tab:reg_res}
\end{table}

\begin{figure}[h] 
    \centering 
    \includegraphics[width=\hsize]{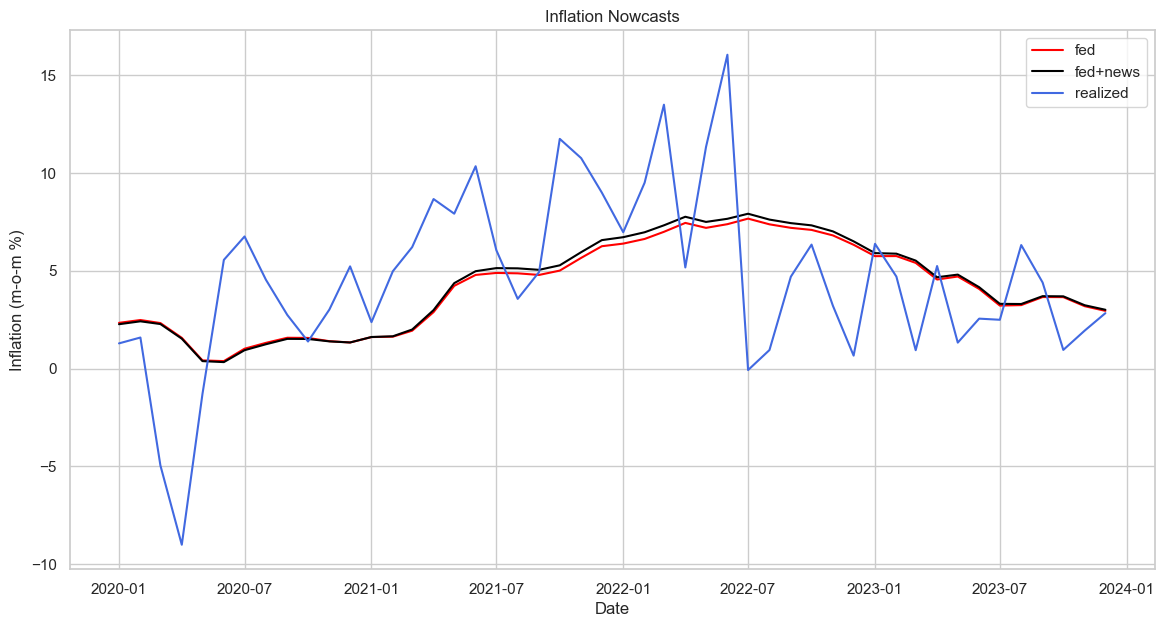} 
    \caption{Annualized Inflation Nowcasts} 
    \label{fig:forecasts} 
\end{figure}

Figure~\ref{fig:forecasts} displays the time series of the annualized models' forecasts. Adding the \texttt{NEWS} index seems to slightly enchance the performance during high-time inflation. Table~\ref{tab:forecasts_res} shows the RMSE and the significance of these forecasts compared to realized inflation. We observe a slight improvement in the forecasts (a reduction of 0.02\% in RMSE), although this improvement is not statistically significant.

\begin{table}[h!]
    \centering
    \begin{tabular}{l c}
        \\[-1.8ex]\hline
        \hline \\[-1.8ex]      
        & \textbf{RMSE}\\
        \midrule
         FED & 0.0409$^{}$ \\
        & (--) \\
         FED+NEWS & 0.0407$^{}$  \\
        & (0.19) \\
        \\[-1.8ex]\hline
        \hline \\[-1.8ex]   
        \textit{Note: } & \multicolumn{1}{r}{$^{*}$p$<$0.1; $^{**}$p$<$0.05; $^{***}$p$<$0.01} \\
    \end{tabular}
    \caption{Inflation Forecasts of the Models. In parenthesis are the p-values of the Giacomini-White test.}
    \label{tab:forecasts_res}
\end{table}

%% file: appendix.tex
\appendix

\section{GPT Labeling Prompting Strategy}
\label{sec:prompt}
This section presents the prompting strategy used to query GPT-4 Turbo API when labeling the dataset.

\begin{tcolorbox}[colback=blue!5, colframe=black!75, width=\columnwidth, boxrule=0.2mm, arc=2mm, auto outer arc, title={Custom Prompting Strategy}]
        Here is a financial news article related to inflation. Please read the article and determine whether it expresses a sentiment that inflation will go up, go down, or if it makes no clear expression. Use any financial knowledge you have.\\
        
        Respond with a \textbf{SINGLE} number (no explanation needed):
        \begin{itemize}
            \item \textbf{1} if the article expresses inflation will go up.
            \item \textbf{-1} if the article expresses inflation will go down.
            \item \textbf{0} if the article sentiment about inflation is neutral.
        \end{itemize}
        
        Here's the article:\\
        \{\textit{ARTICLE}\}
\end{tcolorbox}

\section{Models Cross-Validation}
\label{sec:cv-res}
In this section, we discuss the cross-validation results employed to ascertain the most effective model for analyzing sentiment concerning inflation in news articles. Table~\ref{tab:hyperparams} displays the values of the hyperparameters tested, while Table~\ref{table:results_trad} showcases the test F1-Scores from the best cross-validation setups.

\begin{table}[h]
\centering
\begin{tabular}{p{2.5cm}|p{2.5cm}}
\hline
\multicolumn{2}{c}{\textbf{Logistic Regression}} \\ \hline
\textit{Hyperparameters} & \textit{Values} \\ \hline
penalty & elasticnet \\ \hline
C & 0.01, 0.1, 0.2, 0.5, 1, 5 \\ \hline
l1\_ratio  & 0.0, 0.1, 0.5, 0.7, 1.0 \\ \hline
\multicolumn{2}{c}{\textbf{Random Forest}} \\ \hline
\textit{Hyperparameters} & \textit{Values}\\ \hline
n\_estimators  & 10, 50, 100, 200  \\ \hline
max\_depth  & None, 10, 20, 30 \\ \hline
\multicolumn{2}{c}{\textbf{SVM}} \\ \hline
\textit{Hyperparameters} & \textit{Values} \\ \hline
C  & 0.1, 1, 10 \\ \hline
kernel  & linear, poly, rbf, sigmoid \\ \hline
\multicolumn{2}{c}{\textbf{XGBOOST}} \\ \hline
\textit{Hyperparameters} & \textit{Values} \\ \hline
max\_depth  & 5, 10, 100 \\ \hline
learning\_rate  & 0.01, 0.1, 0.2 \\ \hline
n\_estimators  & 20, 100, 200 \\ \hline
\end{tabular}
\caption{Hyperparameters tested. \label{tab:hyperparams}}
\end{table}

\begin{table*}[h!]
\centering
\begin{tabular}{p{0.3\hsize} p{0.35\hsize} p{0.15\hsize}}
\toprule
\textbf{Model x Embedding} & \textbf{Hyperparameters} & \textbf{F1-Score (Weighted)} \\
\midrule
Logistic Regression + \textbf{W2V} & \{\textbf{'C'}=5, \textbf{'l1\_ratio'}=1.0\}  & 0.6051 \\
Random Forest + \textbf{W2V} & \{\textbf{'max\_depth'}:20, \textbf{'n\_estimators'}:200\}  & 0.7522 \\
SVM + \textbf{W2V} & \{\textbf{'C'}:10, \textbf{'kernel'}: 'rbf'\}  & 0.7571 \\
XGBoost + \textbf{W2V} & \{\textbf{'learning\_rate'}:0.2, \textbf{'max\_depth'}:100, \textbf{'n\_estimators'}: 200\} & \textbf{0.7912} \\
Logistic Regression + \textbf{GloVe} & \{\textbf{'C'}:5, \textbf{'l1\_ratio'}: 0.5,\textbf{'penalty'}: 'elasticnet'\} & 0.5670 \\
Random Forest + \textbf{GloVe} & \{\textbf{'max\_depth'}:None, \textbf{'n\_estimators'}: 200\}  & 0.6876 \\
SVM + \textbf{GloVe} & \{\textbf{'C'}:10, \textbf{'kernel'}: 'poly'\} & 0.5993 \\
XGBoost + \textbf{GloVe} & \{\textbf{'learning\_rate'}:0.1, \textbf{'max\_depth'}:10, \textbf{'n\_estimators'}: 200\} & \textbf{0.7347} \\
\bottomrule
\end{tabular}
\caption{Comparison of different models and embedding techniques using GridSearch with 3-fold Cross Validation}
\label{table:results_trad}
\end{table*}

\section{Full Regression Results}
\label{sec:full_reg}
This section presents additional models result. The models are a combination of the existing Cleveland model regressors with the \texttt{NEWS} index. Table~\ref{tab:full_reg}
present the models regression results.
\begin{table*}[!htbp] \centering
\begin{tabular}{@{\extracolsep{5pt}}lccccc}
\\[-1.8ex]\hline
\hline \\[-1.8ex]
& \multicolumn{5}{c}{\textit{Dependent variable: CPI}} \
\cr \cline{2-6}
\\[-1.8ex] & \multicolumn{1}{c}{fed} & \multicolumn{1}{c}{news} & \multicolumn{1}{c}{fed+news} & \multicolumn{1}{c}{fed-gas+news} & \multicolumn{1}{c}{ccpi+news}  \\
\\[-1.8ex] & (1) & (2) & (3) & (4) & (5) \\
\hline \\[-1.8ex]
 const & 0.021$^{}$ & 0.144$^{***}$ & 0.017$^{}$ & 0.012$^{}$ & 0.003$^{}$ \\
& (0.026) & (0.024) & (0.025) & (0.073) & (0.071) \\
 pi-CCPI & 0.616$^{***}$ & & 0.634$^{***}$ & 0.814$^{**}$ & 0.810$^{**}$ \\
& (0.135) & & (0.133) & (0.383) & (0.381) \\
 pi-FCPI & 0.186$^{***}$ & & 0.176$^{***}$ & -0.105$^{}$ & \\
& (0.064) & & (0.063) & (0.177) & \\
 pi-Gasoline & 0.035$^{***}$ & & 0.034$^{***}$ & & \\
& (0.002) & & (0.002) & & \\
 pi-NEWS & & 0.419$^{*}$ & 0.149$^{*}$ & 0.456$^{*}$ & 0.448$^{*}$ \\
& & (0.237) & (0.081) & (0.232) & (0.230) \\
\hline \\[-1.8ex]
 Observations & 60 & 60 & 60 & 60 & 60 \\
 $R^2$ & 0.892 & 0.051 & 0.898 & 0.126 & 0.121 \\
 Adjusted $R^2$ & 0.886 & 0.035 & 0.890 & 0.080 & 0.090 \\
 Residual Std. Error & 0.064 & 0.185 & 0.062 & 0.181 & 0.180 \\
 F Statistic & 153.612$^{***}$ & 3.131$^{*}$ & 120.877$^{***}$ & 2.702$^{*}$ & 3.921$^{**}$ \\
\hline
\hline \\[-1.8ex]
\textit{Note:} & \multicolumn{5}{r}{$^{*}$p$<$0.1; $^{**}$p$<$0.05; $^{***}$p$<$0.01} \\
\end{tabular}
\caption{Regression Results}
\label{tab:full_reg}
\end{table*}